\documentclass[11pt]{article}
\usepackage{graphicx}

\newcommand{\BABARPubYear}    {04}

\newcommand{\BABARConfNumber} {01}
\newcommand{\SLACPubNumber} {10625}

\input pubboard/babarsym

\long\def\inst#1{\par\nobreak\kern 4pt\nobreak
    {\it #1}\par\vskip 10pt plus 3pt minus 3pt}

\begin{document}
{\pagestyle{empty}

\begin{flushright}
\babar-CONF-\BABARPubYear/\BABARConfNumber \\
SLAC-PUB-\SLACPubNumber \\
August 2004 \\
\end{flushright}

\par\vskip 5cm

\begin{center}
\Large \bf 
A SEARCH FOR THE $\Theta^{*++}$ PENTAQUARK IN $B^{\pm}\rightarrow p \bar{p} K^{\pm}$
\end{center}
\bigskip

\begin{center}
\large The \babar\ Collaboration\\
\mbox{ }\\
\today
\end{center}
\bigskip \bigskip

\begin{center}
\large \bf Abstract
\end{center}
We report the results of a search for the $\Theta^{*++}$ pentaquark
in the decay $B^{+}\rightarrow \Theta^{*++} \bar{p}$ where $\Theta^{*++}\rightarrow pK^+$
using 81$\,\mbox{fb}\,^{-1}$ of data 
collected on the $\Upsilon$(4S) resonance with the \babar\ detector at PEP-II. 
We find an upper limit on the branching fraction of 
$B^{+}\rightarrow \Theta^{*++} \bar{p}$ where $\Theta^{*++}\rightarrow pK^+$
to be $1.5\times 10^{-7}$ for $1.43<m(\Theta^{*++})<1.85\,$GeV/c$^2$,  
$2.4\times 10^{-7}$ for $1.85<m(\Theta^{*++})<2.00\,\mbox{GeV/c}^2$
and  $3.3\times 10^{-7}$ for $2.00<m(\Theta^{*++})<2.36\,\mbox{GeV/c}^2$, at 90$\%$ confidence level.
All results are preliminary.
\vfill
\begin{center}

Submitted to the 32$^{\rm nd}$ International Conference on High-Energy Physics, ICHEP 04,\\
16 August---22 August 2004, Beijing, China

\end{center}

\vspace{1.0cm}
\begin{center}
{\em Stanford Linear Accelerator Center, Stanford University, 
Stanford, CA 94309} \\ \vspace{0.1cm}\hrule\vspace{0.1cm}
Work supported in part by Department of Energy contract DE-AC03-76SF00515.
\end{center}

\newpage
} 

\begin{center}
\small

The \babar\ Collaboration,
\bigskip

%
B.~Aubert,
R.~Barate,
D.~Boutigny,
F.~Couderc,
J.-M.~Gaillard,
A.~Hicheur,
Y.~Karyotakis,
J.~P.~Lees,
V.~Tisserand,
A.~Zghiche
\inst{Laboratoire de Physique des Particules, F-74941 Annecy-le-Vieux, France }
% ...

\end{center}\newpage

\section{INTRODUCTION}
\label{sec:Introduction}
Recently several experimental groups have reported observations of a new, manifestly exotic baryon resonance, 
called the $\Theta^+$(1540) \cite{theta},
with an unusually narrow width ($\Gamma <8\,\mbox{MeV/c}^2$). These results have prompted
a surge of pentaquark searches in experimental data of many kinds \cite{search}. In this paper we will concentrate on the 
exclusive search for pentaquarks in the decay of $B$ mesons. Following the observation of the 
decay $B^+\rightarrow p\bar{p}K^+$\footnote{Charge-conjugates are assumed throughout the paper.} \cite{bad,bad1} 
it was suggested that this decay might include events of the form
$B^+\rightarrow \Theta^{*++}\bar{p}$ where $\Theta^{*++}$ is an $I=1$, $I_3=1$ pentaquark \cite{Browder}.
$\Theta^{*++}$ would be a member of the baryon 27-plet with 
quark content $uuud\bar{s}$. It has been predicted to lie in the region $1.43-1.70\,\mbox{GeV/c}^2$
in the $pK^+$ invariant mass of $B^+\rightarrow p\bar{p}K^+$ candidates
and to have a width of $37-80\,\mbox{MeV}$ \cite{theory}. 
The $pK^+$ cross section is nearly purely elastic in the region of interest 
so a resonance would follow the Breit-Wigner form, with a peak cross section
of about $25\,$mb if the resonance is at $1.7\,\mbox{GeV/c}^2$ and even larger if the mass is
less.  The cross section is measured to be about $12\,$mb at center-of-mass energies spaced by
about $15\,$MeV \cite{PDG}, so its width would need to be considerably less than $15\,$MeV to have
escaped detection. Our limits will not depend on the 
$\Theta^{*++}$ width in any significant fashion. We will search for $\Theta^{*++}$ in the mass region up to 
$2.36\,\mbox{GeV/c}^2$.

\section{THE \babar\ DETECTOR AND DATASET}
\label{sec:babar}
We use data collected on the $\Upsilon$(4S) resonance with the \babar\ detector at \pep2\ to search for $\Theta^{*++}$.
The data sample contains 89 million $B\bar{B}$ pairs, corresponding to an 
integrated luminosity of $81\,\mbox{fb}\,^{-1}$ on the $\Upsilon$(4S) resonance. An additional 
$9\,\mbox{fb}^{-1}$ of data, collected
$40\,\mbox{MeV}$ below the resonance peak (referred to as off-peak data), are used to study the background
from light-quark and $c\bar{c}$ production.

A detailed description of the \babar\ detector can be found elsewhere \cite{nim}; only detector components
relevant to this analysis are mentioned here. Charged-particle trajectories are measured by a five-layer 
double-sided silicon vertex tracker (SVT) and a 40-layer drift-chamber (DCH), operating in the magnetic field 
of a 1.5-T solenoid. Charged particles are identified by combining the measurements of ionization 
energy loss ($dE/dx$) in the DCH and SVT with angular information from a detector of internally reflected
Cherenkov light (DIRC). Photons are identified as isolated electromagnetic showers in a CsI(Tl)
electromagnetic calorimeter. 

\section{ANALYSIS METHOD}
\label{sec:Analysis}

We require that charged tracks have a minimum transverse momentum ($p_{T}$) of $0.1\,$GeV/c, 
at least 12 hits in the DCH, and that they originate from the interaction region 
point within $10\,$cm along the beam direction and $1.5\,$cm in the transverse plane. 

The kaon and proton particle identification is based on $dE/dx$  information from 
the DCH and SVT for $p_{T}<0.7\,$GeV/c or the measured Cherenkov angle and the number of photons observed in 
the DIRC for $p_{T}>0.7\,$GeV/c. 

The $B$ candidate is formed from the proton, the anti-proton and the kaon candidates.
Two kinematic variables are used to isolate the $B^+\rightarrow p\bar p K^+$ signal taking advantage of the kinematic 
constraints of $B$ mesons produced at the $\Upsilon(4S)$. The first is 
the beam-energy-substituted mass, $m_{ES}=[(E^2_{CM}/2+\mbox{\bf p}_i\cdot \mbox{\bf p}_B)^2/E^2_i-{\mbox{\bf p}^2_B}]^{1/2}$, 
where $E_{CM}$ is the total center-of-mass energy of the $e^+e^-$ collision.
Here, the four-momentum of the initial  $e^+e^-$ system is ($E_i, \mbox{\bf p}_i$) and 
${\bf p}_B$ is the momentum of the reconstructed $B$ candidate, both measured in the laboratory frame.
The second variable is $\Delta E=E^*_B-E_{CM}/2$, where 
$E^*_B$ is the $B$-candidate energy in the center-of-mass frame.

Several topological variables provide discrimination between the large continuum background ($e^+e^-\rightarrow q\bar{q},$
where $q=u,d,s,c$), which tends to be collimated along the original quark direction, and more
spherical $B\bar{B}$ events. In order to suppress the dominant continuum background we use a linear combination 
(a Fisher discriminant) of the following four event-shape variables:
cos$\theta^B_{thr}$, the cosine of the angle between the thrust axis of the reconstructed $B$ and the beam axis 
in the center-of-mass frame; cos$\theta^B_{mom}$, the cosine of the angle between the momentum of the reconstructed $B$ 
and the beam axis in the center-of-mass frame; and  zeroth- and second-order Legendre polynomial momentum moments,
$L_0=\sum_{i} |p^*_i|$ and $L_2=\sum_{i} |p^*_i|[(3\cos^2\theta_{thr_{B,i}}-1)/2]$, 
where $p^*_i$ are the center-of-mass momenta for the tracks 
and neutral clusters that are not associated with the $B$ candidate, and $\theta_{thr_{B,i}}$
are the angles between  $p^*_i$ and the thrust axis of the $B$ candidate. 
We optimize the Fisher discriminant coefficients for the best background and $B^+\rightarrow p\bar p K^+$ signal separation using
off-resonance data and $B^+\rightarrow p\bar p K^+$ simulated events that are distributed uniformly in 
phase-space ($B^+\rightarrow p\bar p K^+$ signal Monte Carlo). These event topology requirements retain $67\%$
of $B^+\rightarrow p\bar p K^+$ signal while removing $94\%$ of continuum background. 
We expect $94\%$ of the combinatoric background to come from continuum events and the remaining $6\%$ from $B\bar{B}$ events.

The Fisher discriminant, $m_{ES}$, and $\Delta E$ cuts are optimized to maximize the statistical
sensitivity of the $B^+\rightarrow p\bar{p}K^+$ signal, defined as $S/\sqrt{S+B}$, 
with $S$ and $B$ being estimated numbers of $B^+\rightarrow p\bar p K^+$ signal and background yields in the
Monte Carlo simulation respectively. We assumed the $B^+\rightarrow p\bar p K^+$ signal branching fraction 
of $(5.66^{+0.67}_{-0.57}\pm 0.62)\times 10^{-6}$ \cite{bad} in the optimization.
The $B^+\rightarrow p\bar p K^+$ signal region is defined to be $5.276<m_{ES}<5.286\,$GeV/c$^2$ and  $|\Delta E|<0.029\,$GeV
(signal box) and the $m_{ES}$ sideband region is taken to be $5.20<m_{ES}<5.26\,$GeV/c$^2$ and $|\Delta E|<0.029\,$GeV
for the combinatoric background studies.

The main source of $B\bar{B}$ backgrounds is the $b\rightarrow c\bar cs$ transitions, where 
$B^+\rightarrow X_{c\bar{c}}K^+$, $X_{c\bar{c}}\rightarrow p\bar{p}$ and $X_{c\bar{c}} = \eta_c,\,J/\psi,\,\psi(2S),\,\chi_{c0,1,2}$
(so-called ``charmonium background''). We expect $72\pm 10$ events of this type in the signal box region.
 To check for additional $B\bar{B}$ backgrounds that might
peak in the $B^+\rightarrow p\bar p K^+$ signal region, we study generic $B\bar{B}$ Monte Carlo as
well as a set of samples of exclusive $B$ decay simulated events for potential charmoniumless backgrounds. 
The expected $B\bar{B}$ ``charmoniumless'' background contribution is less than one event in the signal box region. 

\section{SYSTEMATIC STUDIES}
\label{sec:Systematics}
\begin{table}\caption{Systematic uncertainties for the branching fraction of 
$B^+\rightarrow \Theta^{*++}(pK^+)\bar{p}$ without(with) background subtraction.}
\label{systematics}
\begin{center}
\begin{tabular}{||c|c||}
\hline\hline
{Type}&{$\%$ BF}\\\hline
{$B$-counting}&{1.1}\\
{Tracking}&{2.4}\\
{PID}&{6.0}\\
{Event Shape}&{2.0}\\
{Signal Box Cut}&{2.5}\\
{Monte Carlo Statistics}&{1.1}\\
{Background subtraction}&{(1.1)}\\\hline
{Total}&{7.4(7.5)}\\\hline\hline
\end{tabular}
\end{center}
\end{table}

Systematic uncertainties in the analysis are described below and are summarized in Table \ref{systematics}.

The $B^+\rightarrow \Theta^{*++}(pK^+)\bar{p}$ signal efficiency is computed with $B^+\rightarrow p\bar{p}K^+$ simulated events, 
reconstructed and selected using the same procedure as for the data. 
We apply small corrections determined from data to the efficiency calculation to account 
for the overestimation of the tracking and particle-identification systems performance. 
The resulting efficiency, as a function of $m_{pK^+}$, is shown in Fig. \ref{ppk-eff}.
A systematic uncertainty is assigned to each correction to account for the limited size and purity of
the control sample used in computing that correction. For example, for the kaon 
identification, we correct the simulation using a pure sample of $D^{*+}\rightarrow\pi^+D^0$
decays with $D^0\rightarrow K^-\pi^+$ and for the proton identification we use a sample
of $\Lambda\rightarrow p\pi^-$. Conservatively, we take the size of the correction applied as
our systematic error. 
 
In addition, after all the corrections, we compare our $B^+\rightarrow p\bar p K^+$ signal simulation to a
control sample with similar kinematics and final state topology ($B^+\rightarrow J/\psi(e^+e^-)K^+$), 
in order to quantify the ability of the simulation to model the kinematic and event-shape 
variables used in the event selection. The small residual differences in the efficiencies at the cut 
value are assigned as systematic uncertainties affecting the selection procedure.  

For the calculation of the systematic error due to the background subtraction we decrease the $B$-backgrounds by 
the uncertainties in their branching fractions \cite{PDG}. The change in the upper limit for the new background estimation
is $1.1\%$ which we take as a systematic error. 

The systematic error also comprises the uncertainties from the determination of the number of $B\bar{B}$
pairs. We assume that the branching fraction of the $\Upsilon(4S)$ into 
$B\bar{B}$ is $100\%$, with an equal admixture of charged and neutral $B$ final states. We do not include 
any additional uncertainty due to these assumptions. 

\section{RESULTS}
\label{sec:Physics}

\begin{figure}
\begin{center}
\includegraphics[height=7cm, width=12cm]{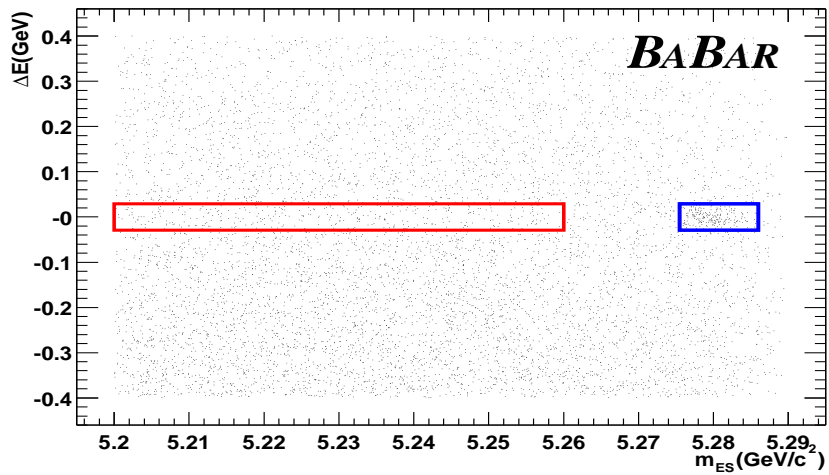}  
  \caption{$m_{ES}\,-\,\Delta E$ distribution of on-peak data 
reconstructed in the $B^+\rightarrow p\bar{p}K^+$ mode. The small box (blue) is ``signal'' box: $5.276<m_{ES}<5.286\,$GeV/c$^2$ and $|\Delta E|<29\,$MeV; 
and the large box (red) is ``sideband'':  $5.20<m_{ES}<5.26\,$GeV/c$^2$ and $|\Delta E|<29\,$MeV.}
  \label{ppk-plane}
\end{center}
\end{figure}

The distribution of events in the $m_{ES}-\Delta E$ plane is shown in Fig. \ref{ppk-plane}. 
We see 212 events in the signal box region. To extract the number of $B^+\rightarrow p\bar p K^+$ signal events 
we loosen signal box cuts on 
$\Delta E$ and fit the $\Delta E$ projection for $5.276<m_{ES}<5.286\,$GeV/c$^2$ and $|\Delta E|<400\,$MeV with a single-Gaussian
distribution for $B^+\rightarrow p\bar p K^+$ signal and a first-order polynomial for the background. 
From that fit we estimate that the 212 total events comprise 40$\pm$2 combinatoric background events and 
188$\pm$17 $B^+\rightarrow p\bar p K^+$ signal events,
including 68$\pm$10 events originating from charmonium decays to $p\bar{p}$. 

\begin{figure}
\begin{center}
\includegraphics[height=18cm,width=9cm]{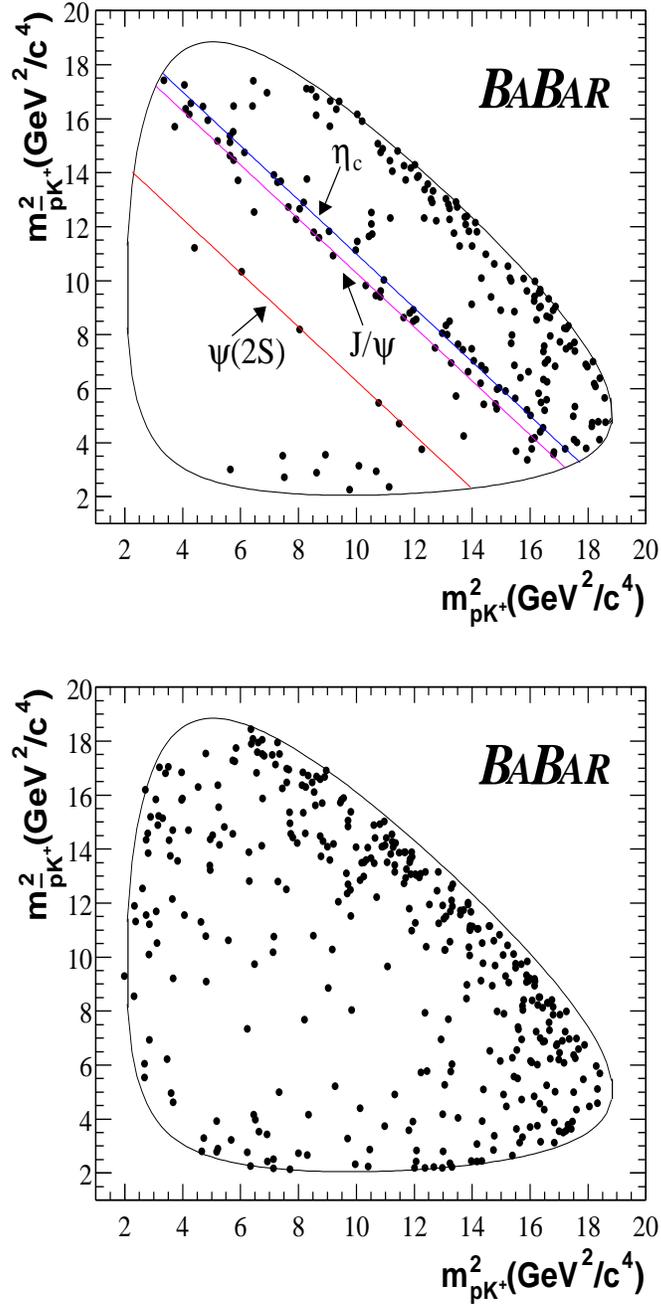}  
  \caption{Dalitz plot of on-peak data reconstructed in $B^+\rightarrow p\bar{p}K^+$ mode. Events in the signal box region (top), 
sideband region (bottom). Note that these distributions are not efficiency-corrected.}
  \label{ppk-dal}
\end{center}
\end{figure}

The Dalitz plots for the events in the signal box (212 events) and the sideband region (368 events) are shown in
Fig. \ref{ppk-dal}. Note that the relative phase-space, or the fraction of Fig. \ref{ppk-dal}(bottom)
in the signal box, is 0.104. The distributions in Fig. \ref{ppk-dal} are not efficiency-corrected.
The Dalitz plot for the events in the signal box (Fig. \ref{ppk-dal}) 
shows a threshold enhancement in the $p\bar{p}$ mass spectrum, as well as three clear bands corresponding to 
$\eta_c$, $J/\psi$ and $\psi(2S)$ events. The background events tend to lie on the edges of the Dalitz plot
because they are dominated by inclusion of random soft tracks.

\begin{figure}
\begin{center}
\includegraphics[height=8cm,width=16cm]{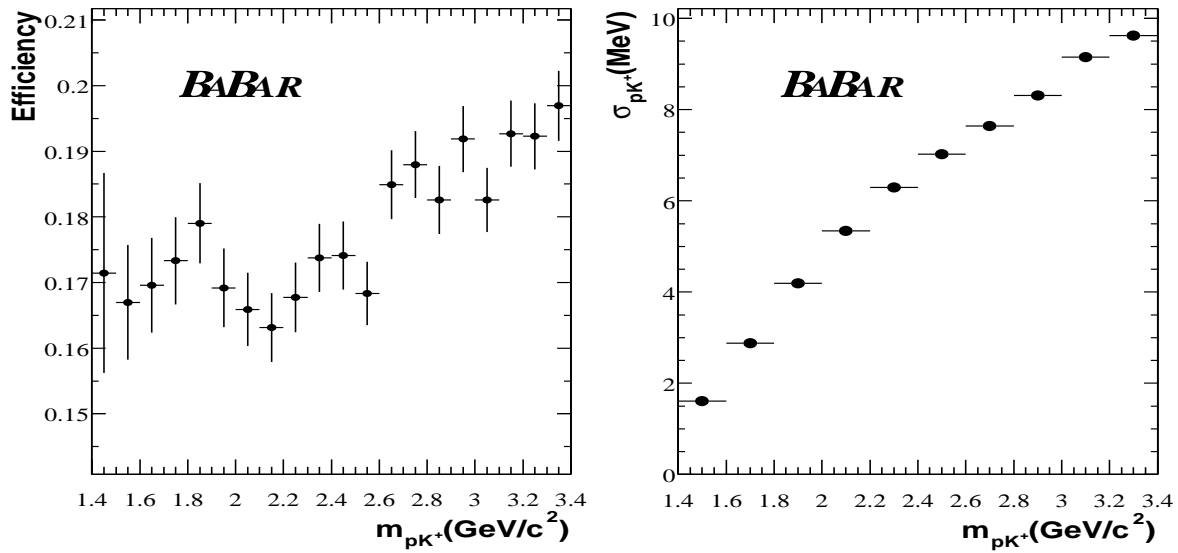}  
\end{center}
\vspace{-1cm}
  \caption{The $B^+\rightarrow p\bar p K^+$ signal reconstruction efficiency(left) and the detector resolution(right) as functions of $m_{pK^+}$}
  \label{ppk-eff}
\end{figure}

\begin{figure}
\begin{center}
\includegraphics[height=8cm,width=12cm]{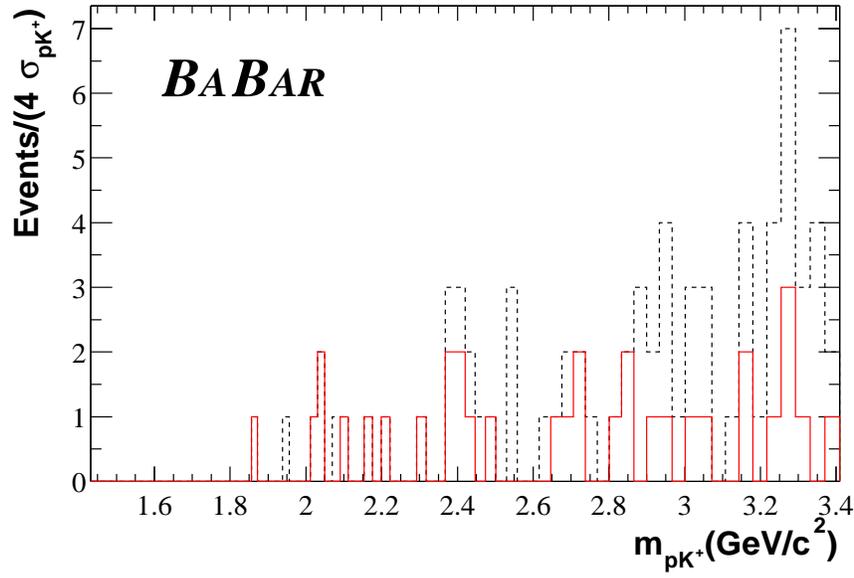}  
  \caption{The $m_{pK^+}$ distribution for data events in $B^+\rightarrow p\bar{p}K^+$ signal box:
events in the charmonium region $2.85<m_{p\bar{p}}<3.15\,$GeV/c$^2$ (solid), events outside the charmonium region (dashed). 
Note that these distributions are not efficiency-corrected.}
  \label{ppk-theta}
\end{center}
\end{figure}

\begin{figure}
\begin{center}
\includegraphics[height=8cm,width=12cm]{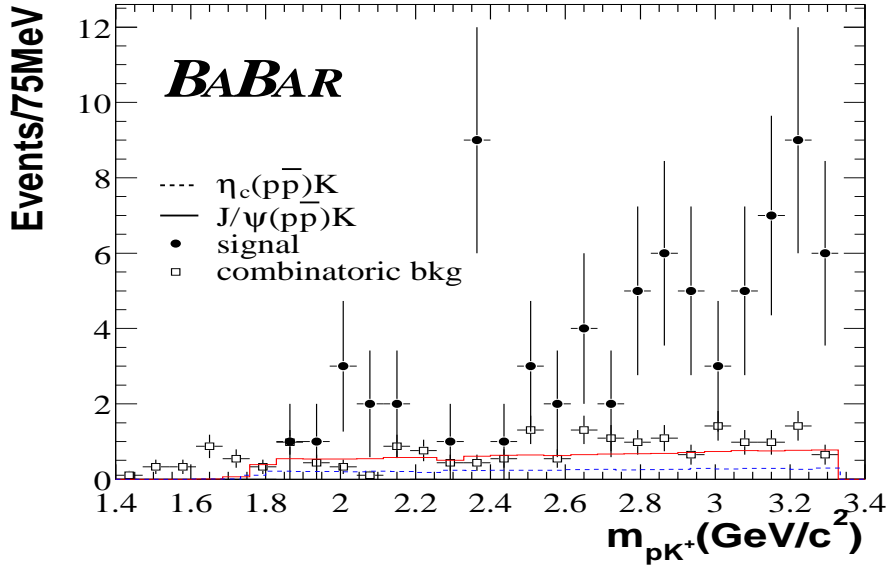}  
  \caption{The $m_{pK^+}$ distributions for data reconstructed as  $B^+\rightarrow p\bar{p}K^+$ (dots), 
$m_{ES}$ sideband (empty squares) and, Monte Carlo, $B^+\rightarrow \eta_c(p\bar{p})K^+$(dashed line) 
and $B^+\rightarrow J/\psi(p\bar{p})K^+$(solid line). Note that these distributions are not efficiency-corrected.}
  \label{ppk-thetabkg}
\end{center}
\end{figure}

\begin{figure}
\begin{center}
\includegraphics[height=8cm,width=12cm]{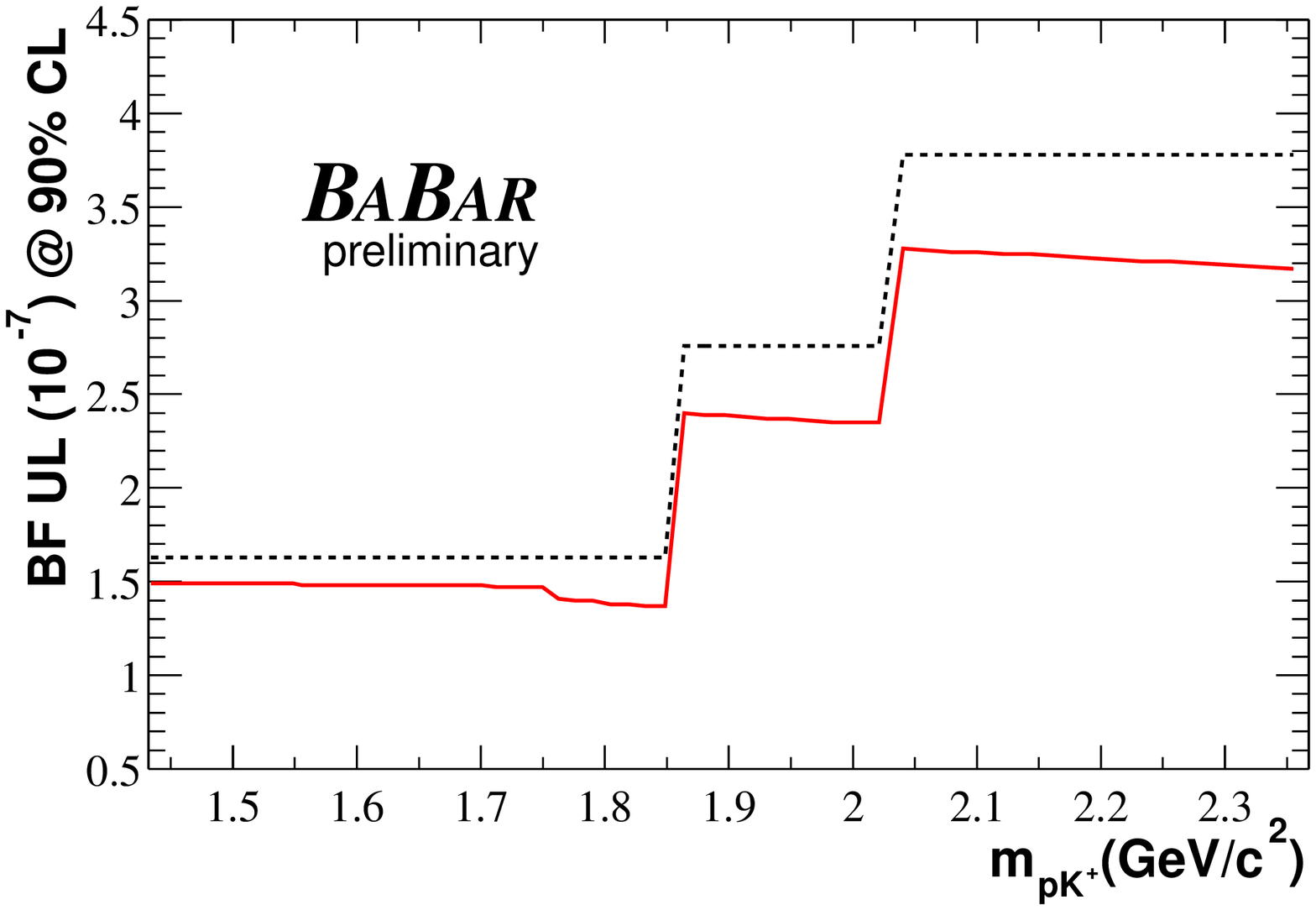}  
\end{center}
  \caption{Upper Limit on the branching fraction of $B^+\rightarrow \Theta^{*++}(pK^+)\bar{p}$
at 90$\%$ confidence level with the assumption of no background(dashed), with background as determined from
data and Monte Carlo (solid). The systematic error correction is included in the limits.}
\label{ppk-ul}
\end{figure}

As we are interested only in the low $m_{pK^+}$ region the following figures will be limited to $m_{pK^+}$ up to 
$3.4\,\mbox{GeV/c}^2$ or the total of 75 events in the signal box region. 
It is convenient to represent data in two different ways: in Fig. \ref{ppk-theta} we 
separate the events into those inside the charmonium window and those outside, where as in Fig. \ref{ppk-thetabkg}
we emphasize the different background contributions to the data.

We search for $\Theta^{*++}$ pentaquark in the $pK^+$ mass spectrum, shown in Fig. \ref{ppk-theta}. 
The binning corresponds to $4\cdot\sigma_{pK^+}$, where $\sigma_{pK^+}$ is the detector resolution shown in Fig. \ref{ppk-eff}(right).
The average $B^+\rightarrow \Theta^{*++}(pK^+)\bar{p}$ signal efficiency is $(17.0\pm 0.2)\%$ for $1.43<m_{pK^+}<2.40\,$GeV/c$^2$. 
We observe no events for $m_{pK^+}<1.85\,$GeV/c$^2$. 

The background contributions are shown in Fig. \ref{ppk-thetabkg}. The $m_{pK^+}$ distribution of
the combinatoric background is obtained from the events in the data $m_{ES}$ sideband region and 
is scaled to the expected number of the combinatoric background events in the signal box. 
The shape and amount of $B^+\rightarrow\eta_c(p\bar{p})K^+$ and $B^+\rightarrow J/\psi(p\bar{p})K^+$ background contributions
are determined from the simulation and scaled by their respective branching fractions \cite{PDG}.
There is no contribution to the background from $B^+\rightarrow\eta_c(p\bar{p})K^+$ for $m_{pK^+}<1.80\,$GeV/c$^2$ and
$B^+\rightarrow J/\psi(p\bar{p})K^+$ for $m_{pK^+}<1.75\,$GeV/c$^2$.

To set an upper limit at 90$\%$ confidence level on the branching fraction
of $B^+\rightarrow \Theta^{*++}(pK^+)\bar{p}$ we count events in each of the $m_{pK^+}$ mass bins in Fig. \ref{ppk-theta}
assuming that all the events observed are $B^+\rightarrow \Theta^{*++}(pK^+)\bar{p}$ signal events.
To simplify the presentation of the upper limit on the branching fraction as a function $m_{pK^+}$
we assume that number of events in each of the bins in $m_{pK^+}$ is equal to the maximum number of events per bin for each of the
$m_{pK^+}$ regions (see Table \ref{results}). 

\begin{table}\caption{The upper limit for the branching fraction of 
$B^+\rightarrow \Theta^{*++}(pK^+)\bar{p}$ as a function $m_{pK^+}$ without(with) background subtraction.}
\label{results}
\begin{center}
\begin{tabular}{||c|c|c|c||}
\hline\hline
{Mass Region,}&{Maximum Events Observed}&{BF UL ($10^{-7}$) @}&{BF UL ($10^{-7}$) @}\\
{GeV/c$^2$}&{in any $m_{pK^+}$ bin}&{90$\%$ CL without bkg}&{90$\%$ CL with bkg}\\\hline
{1.43$<m_{pK^+}<$1.85}&{0}&{1.63}&{1.49}\\
{1.85$<m_{pK^+}<$2.00}&{1}&{2.76}&{2.40}\\
{2.00$<m_{pK^+}<$2.36}&{2}&{3.78}&{3.28}\\\hline\hline
\end{tabular}
\end{center}
\end{table}

We use two methods to determine the upper limit. In the first one we assume that there is no background contribution. 
We calculate from Table 31.3 \cite{PDG} the Bayesian upper limit for 90$\%$ confidence level as a function of $m_{pK^+}$
assuming Poisson-distributed events in the absence of background.
The resulting values are shown in Fig. \ref{ppk-ul} and given in Table \ref{results}. To account for systematic errors we increase the upper limit by the total 
systematic error ($7.5\%$).

To calculate the upper limit in the presence of background we use a tool described in \cite{calcul}. It uses
toy Monte Carlo technique to calculate an upper limit in presence of uncertainties on the efficiency and the number of 
expected background events. We assume all the systematic errors but the systematics on background and $B$-counting to contribute 
to the uncertainty on the efficiency ($7.3\%$). To estimate the number of expected background events we fit a first-order
polynomial to the $pK^+$ mass spectrum of the combinatoric background events as well as 
$B^+\rightarrow\eta_c(p\bar{p})K^+$ and $B^+\rightarrow J/\psi(p\bar{p})K^+$ Monte Carlo events (so-called peaking $B$-background).
The uncertainty on the background comes from the statistical error on the fit as well as the systematic error on the background.
The resulting values of the upper limit as a function of $m_{pK^+}$ increased by the systematic
error on $B$-counting ($1.1\%$) are given in Table \ref{results} and shown in Fig. \ref{ppk-ul}.

\section{SUMMARY}
\label{sec:Summary}
Using 81$\,\mbox{fb}^{-1}$ of on-peak data accumulated by the \babar\ detector, 
we set an upper limit at 90$\%$ confidence level on the branching fraction
of $B^+\rightarrow \Theta^{*++}(pK^+)\bar{p}$ to be $1.49\times 10^{-7}$
for $1.43<m(\Theta^{*++})<1.85\,$GeV/c$^2$,  $2.40\times 10^{-7}$ for $1.85<m(\Theta^{*++})<2.00\,\mbox{GeV/c}^2$
and  $3.28\times 10^{-7}$ for $2.00<m(\Theta^{*++})<2.36\,\mbox{GeV/c}^2$.

\section{ACKNOWLEDGMENTS}
\label{sec:Acknowledgments}
We are grateful for the 
extraordinary contributions of our \pep2\ colleagues in
achieving the excellent luminosity and machine conditions
that have made this work possible.
The success of this project also relies critically on the 
expertise and dedication of the computing organizations that 
support \babar.
The collaborating institutions wish to thank 
SLAC for its support and the kind hospitality extended to them. 
This work is supported by the
US Department of Energy
and National Science Foundation, the
Natural Sciences and Engineering Research Council (Canada),
Institute of High Energy Physics (China), the
Commissariat \`a l'Energie Atomique and
Institut National de Physique Nucl\'eaire et de Physique des Particules
(France), the
Bundesministerium f\"ur Bildung und Forschung and
Deutsche Forschungsgemeinschaft
(Germany), the
Istituto Nazionale di Fisica Nucleare (Italy),
the Foundation for Fundamental Research on Matter (The Netherlands),
the Research Council of Norway, the
Ministry of Science and Technology of the Russian Federation, and the
Particle Physics and Astronomy Research Council (United Kingdom). 
Individuals have received support from 
CONACyT (Mexico),
the A. P. Sloan Foundation, 
the Research Corporation,
and the Alexander von Humboldt Foundation.

\end{document}